\documentclass[final,3p,times,twocolumn]{elsarticle}

\usepackage{graphicx}

\usepackage{amssymb}
\usepackage{amsmath}

\usepackage[colorlinks, citecolor=red]{hyperref}

\biboptions{comma,square,sort&compress}

\begin{document}

\begin{frontmatter}



\title{Transport Measurements on Nano-engineered Two Dimensional Superconducting Wire Networks}


\author{W. J. Zhang} \fnref{label1}
\ead{zhangweijun@ssc.iphy.ac.cn}
\author{S. K. He, H. Xiao, G. M. Xue, Z. C. Wen, X. F. Han, S. P. Zhao, C. Z. Gu, X. G. Qiu}
\address{Beijing National Laboratory for Condensed Matter Physics, Institute of Physics, 

Chinese Academy of Sciences, P. O. Box 603, Beijing 100190, China}

\begin{abstract}
Superconducting triangular Nb wire networks with high normal-state resistance are fabricated by using a negative tone hydrogen silsesquioxane (HSQ) resist. Robust magnetoresistance oscillations are observed up to high magnetic fields and maintained at low temperatures, due to the effective reduction of wire dimensions. Well-defined dips appear at integral and rational values (1/2, 1/3, 1/4) of the reduced flux $f$ $=$ $\Phi/\Phi_0$, which is the first observation in the triangular wire networks. These results are well consistent with theoretical calculations for the reduced critical temperature as a function of $f$.
\end{abstract}


\end{frontmatter}


\section{Introduction}
\label{1}
Wire networks are particularly useful model to investigate the lowest energy spectrum of a lattice for tight binding electrons in a homogeneous magnetic field \cite{18}. To our knowledge, although experimental and theoretical studies have discussed the triangular geometry in various arrays (e.g.$~$fractal Sierpinski gaskets \cite{4}, Josephson junction arrays \cite{3,5,15}, array of antidots \cite{6,7}), a direct experimental result of triangular wire networks has not been reported on magnetoresistance measurements in pervious works. On the other hand, the phase interference phenomena in wire networks are quite sensitive to coherence of the order parameter \cite{1,2}. Thus, it is important to fabricate the wires with cross section dimensions comparable with the temperature dependent coherence length $\xi(T)$ \cite{12}.

Here we investigate the triangular wire networks, fabricated by a negative tone resist hydrogen silsesquioxane (HSQ, Dow Corning Co.). HSQ is a high sensitive and etching durable resist, which can effectively reduce the dimensions of wires. Due to the weak localization, our samples made of pure Nb thin films show a high normal-state resistance, in contrast to the works done on low-$T_c$ Al \cite{2}. These samples with a lattice constant $a$ much larger than $\xi(T)$, can be viewed as weakly coupled wire networks \cite{2,16}. Pronounced dips at both integral and fractional reduced flux $f$ $=$ $\Phi/\Phi_0$ are found in the magnetoresistance measurements, which are consistent well with the numerical results. Interestingly, the magnetoresistance oscillations can survive at much high magnetic fields and sustain at relatively low temperatures.


\section{Experimental details}
\label{2}
\begin{figure}[htb]
  \centerline{\includegraphics[width=0.7\columnwidth]{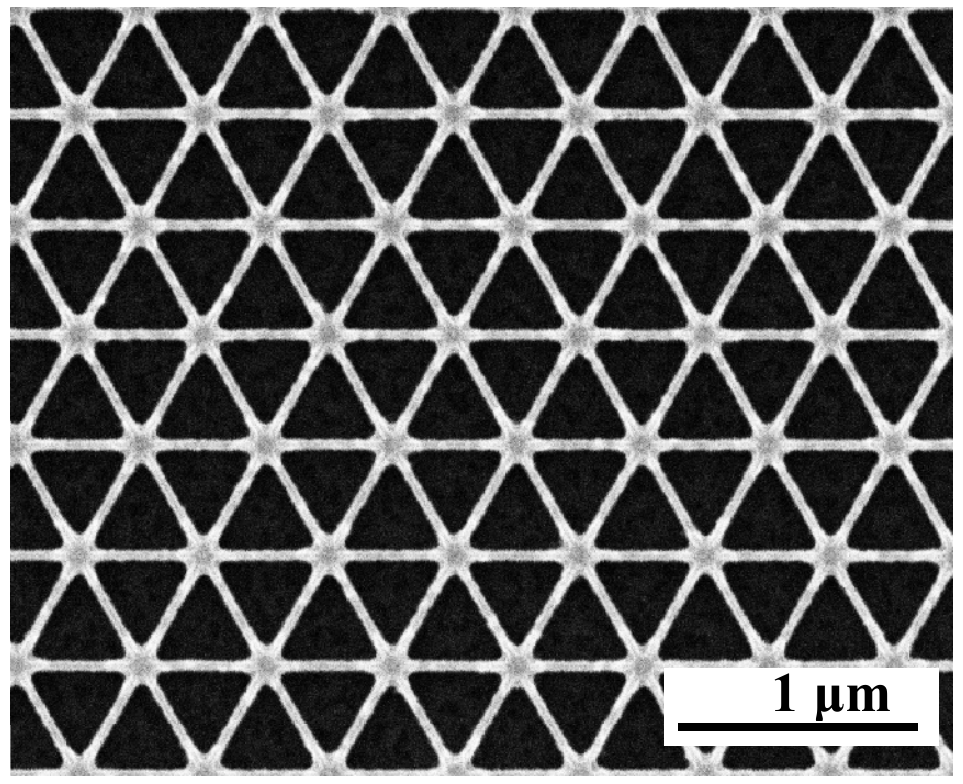}}
  \caption{(color online) Image of scanning electron microscope for a triangular Nb wire network, with a lattice constant $a$ $=$ 525 nm and a wire width $\Delta W$ $=$ 40 nm.
}
  \label{Fig1}
\end{figure}

Nb thin films with a thickness of 50 nm were deposited on oxidized Si substrates by magnetron sputtering. 100 nm thick HSQ resist was spun on the thin film and baked at 170 $^{\circ}$C for 120 seconds. Designed patterns including four-probe microbridges were written by electron beam lithography (EBL) on HSQ resist. The resist was developed for 90 seconds in a 6.25\% tetramethylammonium hydroxide (TMAH) aqueous solution. Finally, the patterns were transferred to the Nb layer by reactive ion etching (RIE) in SF$_{6}$ and O$_{2}$ plasmas. Wire networks were obtained in the centers of the Nb microbridges, with an area of 36 $\mu$m $\times$ 37 $\mu$m. One of the Nb microbridges was intentionally unpatterned and used as a reference. Figure.$~$\ref{Fig1} show a typical image of scanning electron microscope for a triangular Nb wire network, with a lattice constant $a$ $=$ 525 nm and a uniform wire width $\Delta W$ $=$ 40 nm.

Transport measurements have been performed in Physical Properties Measurement System (PPMS-14, Quantum Design Inc.). Phase lock-in amplifiers (SR-830) are used for ac-current supplies with a frequency of 30.9 Hz. Magnetic field is applied perpendicular to the surface of thin films. The temperature stability is approximately 2 mK during the measurements. The superconducting coherence length $\xi(0)$ is 8.6 nm and the penetration depth $\lambda(0)$ is 97.2 nm, determined by measuring the $H(T)$ of the reference Nb microbridge \cite{Tinkham.BOOK.1996}.

\section{Results and Discussions}
\label{3}
\begin{figure}[htb]
  \centerline{\includegraphics[width=0.9\columnwidth]{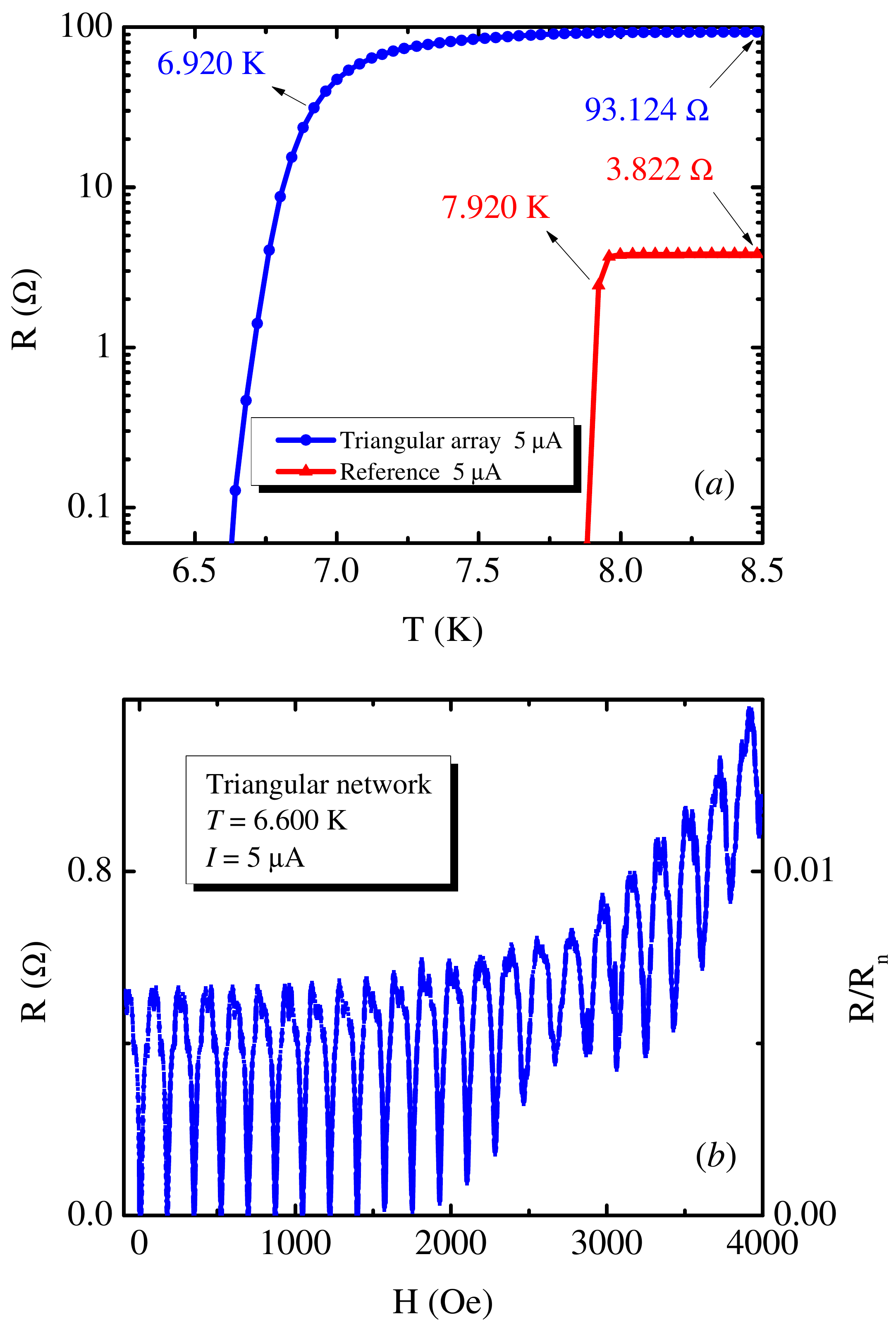}}
  \caption{(color online) (a) Resistance as a function of temperature for triangular wire network and reference Nb film, with $I$ $=$ 5 $\mu$A. (b) Magnetoresistance of the triangular wire network, measured at $T$ $=$ 6.600 K ($t$ $=$ $T/T_c$ $=$ 0.954), with $I$ $=$ 5 $\mu$A. The right axis indicates that the reduce resistance $R/R_n$ is less than 1.5\%.
}
  \label{Fig2}
\end{figure}

As shown in Fig.$~$\ref{Fig2}(a), the reference Nb thin film has a critical temperature $T_{c}$ of 7.920 K and a superconducting transition width $\Delta T_c$ of 63 mK (10\% - 90\%$R_{n}$ criterion, where $R_n$ $=$ 3.822 $\Omega$ is the normal-state resistance at 8.5 K). In contrast to the reference film, broad superconducting transition is visible for the thin film with triangular network. $T_c$ and $\Delta T_c$ of the fabricated triangular network is 6.920 K and 672 mK, respectively. The expansion of $\Delta T_c$ is due to the reduced wire dimensions in the fabrication, as commonly seen in nanowires \cite{10}. Note that $T_c$ is defined by the temperature where $dR/dT$ has the largest value. The resistance of the network is nearly 12 times higher than that of a triangular network with a wider strands. With $R_n$ $=$ 93.124 $\Omega$ at 8.5 K, we can roughly estimate the resistance of a single wire equals to be 107.091 $\Omega$.

Figure \ref{Fig2}(b) plots the resistance $R$ as a function of the magnetic field at $T$ $=$ 6.600 K, with current $I$ $=$ 5 $\mu$A. The maximus of $R$ is found smaller than 1.5\%$R_n$ in the present curve. Non-monotonic magnetoresistance oscillations are observed at least up to 4000 Oe, with a series of dips appearing at periodic values of the magnetic field. At a lower temperature of 6.480 K, the upper critical field of the network is as large as 1.1489 T (50\%$R_n$ criterion), which is much larger than the theoretical bulk critical field $H_{cb}$ $=$ $\Phi/[2\pi\xi(t)^2]$ $=$ 0.2832 T at the same reduced temperature $t$ $=$ $T/T_c$ $=$ 0.936. This indicates a strong enhancement of critical field in the reduced dimensional samples. The oscillation period $H_1$ $=$ 173.9 Oe corresponds to one quantum flux $\Phi_0$ $=$ $hc/2e$ per elementary triangular loop in the network. These oscillations reflect the collective effect of quantization phenomena similar to that of Little-Parks effect \cite{9}. The enclosed area inside elementary loops, estimated from this magnetic period, corresponds to triangular cells of side $a$ $=$ $\sqrt{\frac{4\Phi_0}{\sqrt{3}H_1}}$ $=$ 524 nm. This value is closed to the one extracted from the image of scanning electron microscope (SEM): $a$ = 525 $\pm$ 4 nm. 


Due to thin width of the wire (40 nm) and broad transition width, we observe the magnetoresistance oscillations in a wide range of temperatures. The present curve  in Fig.$~$\ref{Fig2}(b) is measured at $t$ $=$ 0.954, which is lower than the reported typical value ($t$ $=$ 0.990) for perforated samples. Similar phenomena have also been found in TiN thin film with square arrays of antidots \cite{11} and high-$T_c$ NbN films with ferromagnetic nanowires arrays \cite{19}. In the temperature range of $t$ $=$ 0.950 $\sim$ 0.990, $\xi(t)$ varies between 39 nm and 86 nm. 
Thus, at the measured temperatures, $\xi(t)$ is always comparable with the width of thin wire, but much smaller than the lattice constant $a$. Our sample can be treated as a weakly coupled wire network system.
\begin{figure}[htb]
  \centerline{\includegraphics[width=0.9\columnwidth]{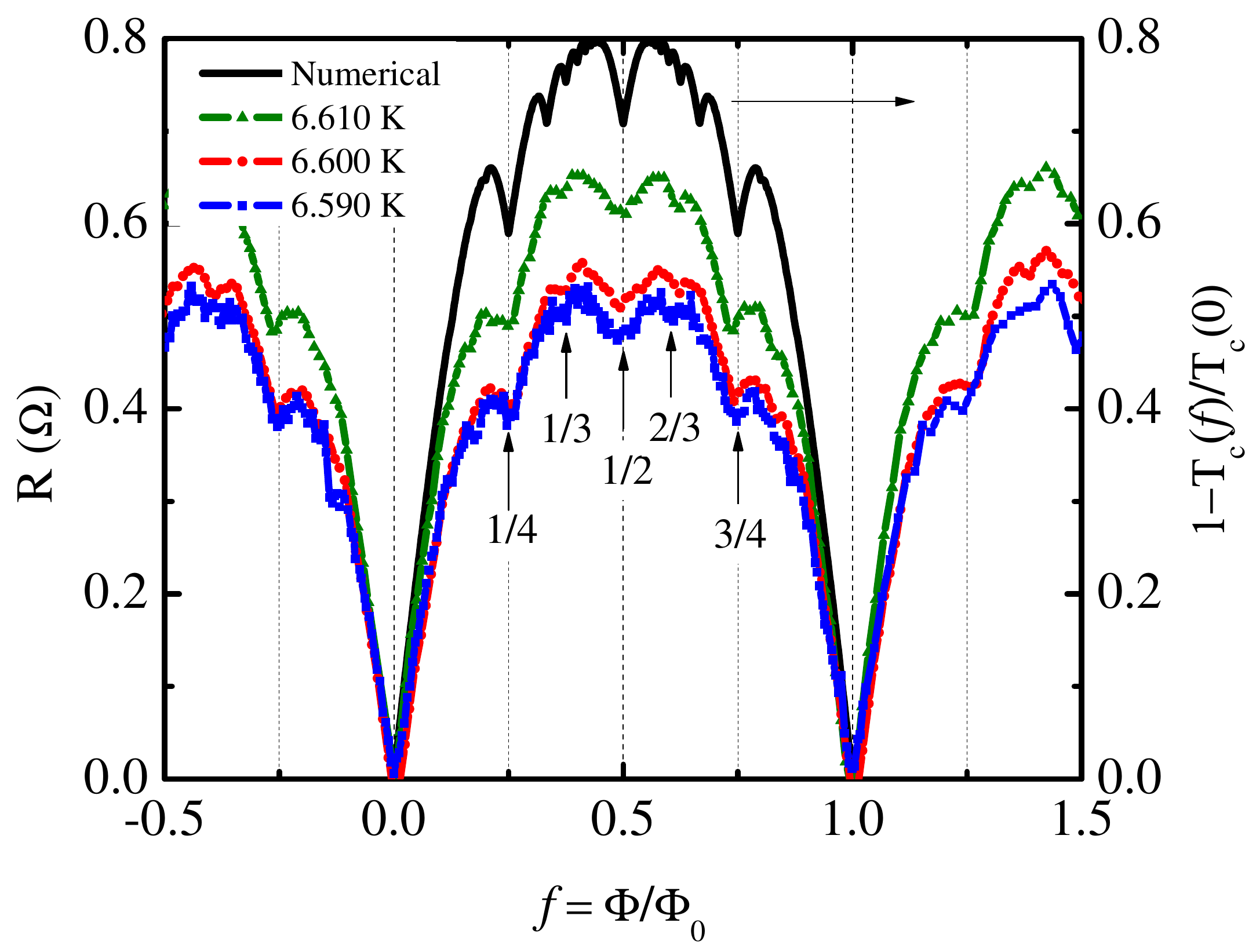}}
  \caption{(color online) Resistance as a function of the reduced flux $\Phi/\Phi_0$, measured at several temperatures. Fractional reduced flux $f$ $=$ $\Phi/\Phi_0$ $=$ 1/4, 1/3, 1/2, 2/3 and 3/4 are observed. Solid curve is the numerical result of the mean field G-L theory, with reduced units [$1 - T_c(f)/T_c(0)$] vs $f$. (see right axis)
}
  \label{Fig3}
\end{figure}

As shown in Fig.$~$\ref{Fig3}, magnetoresistance curves for the same network are measured at several temperatures. Distinct dips at fractional matching fields (1/4, 1/3, 1/2, ... etc.) are visible when one zooms in the low field regime. The dips are symmetrically located around $f$ $=$ 1/2. All these features show a collective behaviors, where stable vortex configuration has been found at certain magnetic field for square wire networks \cite{17}.

The superconducting wire network near the second order phase boundary can be well described by the de Gennes-Alexander formula based on the linearized Ginzburg-Landau theory \cite{12,13}. The equations for wire networks leads to an eigenvalue problem which is expressed in terms of the ground state eigenvalue $\epsilon(f)$ of the tight band spectrum \cite{12,18}. For the triangular lattice, using Landau gauge and periodic boundary conditions, the corresponding equations have the following forms:
\begin{equation}\label{eqx1}
\begin{split}
\epsilon(f) \Delta_{n} &= 2\cos[(2n - 1)\pi f - \frac{k_y}{2}] e^{-i\frac{k_y}{2}}\Delta_{n-1} \\
&+ 2\cos(4n\pi f - k_y)\Delta_{n} \\
&+ 2\cos[(2n + 1)\pi f - \frac{k_y}{2}]e^{i\frac{k_y}{2}}\Delta_{n+1},
\end{split}
\end{equation}
where $\epsilon(f)$ is the eigenvalue equation. $\Delta_{n}$ is the value of order parameter at node $n$. $k_y$ $=$ $2\pi\frac{k-1}{N}$, $k$ $=$ ($1, ...,N$) implies the periodic condition along the y axis.

\begin{eqnarray}\label{eqx2}
1 - \frac{T_c(f)}{T_c(0)} = \frac{\xi(0)^2}{a^2}[\arccos(\frac{\epsilon(f)}{6})]^2,
\end{eqnarray}

The numerical results of Eq.$~$\eqref{eqx2} have been plotted in Fig.$~$\ref{Fig3},  with reduced units $[1 - T_c(f)/T_c(0)]$ as a function of reduced flux $f$. As can be seen, the theoretical curve shows the same fine structures. The location and the relative magnitude of the dips are reproduced at rational values $\Phi/\Phi_0$ $=$ (1/2, 1/3, 1/4, ...) and in a good agreement with the experimental data.
The fine structures of the $T_c(H)$ of triangular wire network have also been studied by using analytical approach based on multiple-loop Aharonov-Bohm Feynman path integrals \cite{15}.
%

%

In conclusion, we have investigated the responds of a superconducting triangular Nb wire network to a perpendicular magnetic field, which shows a high normal-state resistance. Due to the effective reduction of wire dimensions, strong magnetoresistance oscillations have been observed up to much high magnetic fields and maintained at low temperatures. In addition, pronounced dips at integral and fractional reduce flux $f$ $=$ (1/2, 1/3, 1/4) are found, which is in a good agreement with the theoretical analysis. Our samples are promising to study phase fluctuations and superconducting to insulator transition in ultrathin nanowire systems.

\section*{Acknowledgement}
This work is supported by National Basic Research Program of China (No.$~$2009CB929102, 2011CBA00107, 2012CB921302) and National Science Foundation of China (No.$~$10974241, 91121004, 11104335).











\end{document}